\documentclass[]{spie}  
\usepackage[]{graphicx}
\usepackage[]{xspace}

\newcommand{\NACO}{N{\large \bf A}C{\large \bf O}}

\newcommand{\naco}{N{\scriptsize A}C{\scriptsize O}\xspace}
\newcommand{\arcsec}{\mbox{$''$}}
\newcommand{\arcmin}{\mbox{$'$}}

\def\bc{\begin{center}}
\def\ec{\end{center}}
\def\bl{\begin{list}{$\bullet$}{}}
\def\el{\end{list}}

\title{Status and new operation modes of the versatile VLT/\NACO}

\author{Julien H. V. Girard\supit{a}, 
Markus Kasper\supit{b}, 
Sascha P. Quanz\supit{c}, 
Matthew A. Kenworthy\supit{d}, \\
Sridharan Rengaswamy\supit{a},
Rainer Sch\"odel\supit{e},
Alexandre Gallenne\supit{a},
Stefan Gillessen\supit{f},\\
Nicolas Huerta\supit{a}, 
Pierre Kervella\supit{g},
Nick Kornweibel\supit{b},
Rainer Lenzen\supit{h},
Antoine M\'erand\supit{a},
Guillaume Montagnier\supit{a}, 
Jared O'Neal\supit{a}, 
G\'erard Zins\supit{i}
and
the \naco IOT\supit{a,b}
\skiplinehalf
\supit{a}European Southern Observatory (ESO), Casilla 19001, Vitacura, Santiago, Chile;\\
\supit{b}European Southern Observatory, Karl-Schwarzschild-Stra\ss e 2, D-85748 Garching, Germany;\\
\supit{c}Institute of Astronomy, ETH Zurich, CH-8092 Zurich, Switzerland;\\
\supit{d}Leiden Observatory, Leiden University, P.O. Box 9513, 2300 RA, Leiden, The Netherlands;\\
\supit{e}Instituto de Astrof\'{\i}sica de Andaluc\'{\i}a, Camino Bajo de Huetor 50, 18008 Granada, Spain;\\
\supit{f} Max-Planck-Institut f\"ur extraterrestrische Physik, Giessenbachstra\ss e 2, D-85748, Germany;\\
\supit{g}LESIA, Observatoire de Parist, 5 place J. Janssen, 92195 Meudon, France;\\
\supit{h}Max Planck Institut f\"ur Astronomie, K\"onigsthul 17, D-69117 Heidelberg, Germany;\\
\supit{i}LAOG, Universit\'e Joseph Fourier / CNRS, BP 53, F-38041 Grenoble, France;
}

\authorinfo{Corresponding author's e-mail: jgirard@eso.org}
 
\begin{document} 
  \maketitle 

\begin{abstract}
This paper aims at giving an update on the most versatile adaptive optics fed instrument to date, the well known and successful \naco\footnote{\naco: Nasmyth Adaptive Optics System {\bf NA}OS and the Near-Infrared Imager and Spectrograph {\bf CO}NICA}. Although \naco is only scheduled for about two more years\footnote{\naco's decommissioning is currently scheduled at the end of 2011.} at the Very Large Telescope (VLT), it keeps on evolving with additional operation modes bringing original astronomical results. The high contrast imaging community uses it creatively as a test-bench for SPHERE\footnote{SPHERE: Spectro-Polarimetric High-contrast Exoplanet REsearch} 
and other second generation planet imagers. A new visible wavefront sensor (WFS) optimized for Laser Guide Star (LGS) operations has been installed and tested, the cube mode is more and more requested for frame selection on bright sources, a seeing enhancer mode (no tip/tilt correction) is now offered to provide full sky coverage and welcome all kind of extragalactic applications, etc. The Instrument Operations Team (IOT) and Paranal engineers are currently working hard at maintaining the instrument overall performances but also at improving them and offering new capabilities,
providing the community with a well tuned and original instrument for the remaining time it is being used. 
The present contribution delivers a non-exhaustive overview of the
new modes and experiments that have been carried out in the past months.

\end{abstract}



\section{\naco: particularities, science return, context and lifetime}
\label{sec:intro}  

\naco is an adaptive optics (AO) fed near infrared (NIR) imager and spectrometer at the 8-meter Very Large Telescope (VLT)  ran by the European Southern Observatory (ESO). It was commissioned in 2001\cite{brandner2002} and offered to the community in october 2002  (periode 70). For a detailed description of \naco see Lenzen et al. \cite{lenzen2003} and Rousset et al. \cite{rousset2003}. The first reference describes the scientific instrument CONICA and its many operating modes and the second one gives an overview of the NAOS AO system.

NAOS' originality comes from the fact it has several Shack-Hartmann type wavefront sensors (WFS) allowing classical visible light sensing, LGS sensing at 589 nm, and last but not least an near infrared (NIR) WFS allowing to use very red object as natural guide stars (NGS). This latter feature has been a tremendous advantage for the long-term study of the galactic center \cite{schoedel2002, genzel2003, eckart2008} as well as to probe dust embedded young stellar objects \cite{murakawa2008}, IR-bright active galactic nuclei, characterize asteroids and other solar system bodies \cite{sicardy2006}, directly image low-mass stellar companions \cite{chauvin2004, neuhauser2005}, etc. 

Being an almost 10-year old AO system sensing the turbulent wavefront over  the 8-meter VLT primary mirror with up to 185 sub-pupils, NAOS performances in terms of Strehl ratios are modest in today's context. However, the number and combinations of operation modes offered by CONICA and its six different cameras makes the \naco instrument still very attractive. The addition of new modes over the past few years \cite{kasper2005} have made \naco an even more versatile instrument offering a four-quadrant phase mask (4QPM)\cite{rouan2000}, simultaneous differential imaging (SDI) and polarimetric capabilities \cite{witzel2010}.  In the high contrast imaging world, \naco is one of the few AO systems that runs an Aladdin-3 detector (upgraded in 2005) allowing to operate between 1 and 5 $\mu$m. Lately, it has been shown that if shorter wavelengths provide in general better spatial resolution, the 4 $\mu$m region (L-band) is particularly suited for exoplanet search and characterization \cite{kasper2009}. In 2009, the spectrum of the exoplanet HR 8799 c was obtained thanks to \naco grism spectroscopy in the L-band \cite{janson2010}.
 
\section{HEALTH, OPERATIONS \& PERFORMANCES} 
\label{sec:health}

\naco is maintained and operated by ESO staff at the Paranal Observatory in northern Chile. The most common setups are supported in Service Mode (SM) operations so that users do not have to travel to the observatory and they do get the atmospheric constraints  they requested as well as  quality controlled data products \cite{dobrzycka2004}. 
This section focuses onto two modes of operations which can be coupled with many other modes: LGS and cube mode (detector readout mode). They allow different observing strategies and now influence the results, the efficiency  and therefore the capabilities of \naco.

\begin{figure}[h!]
\bc
\includegraphics[width=0.85\textwidth]{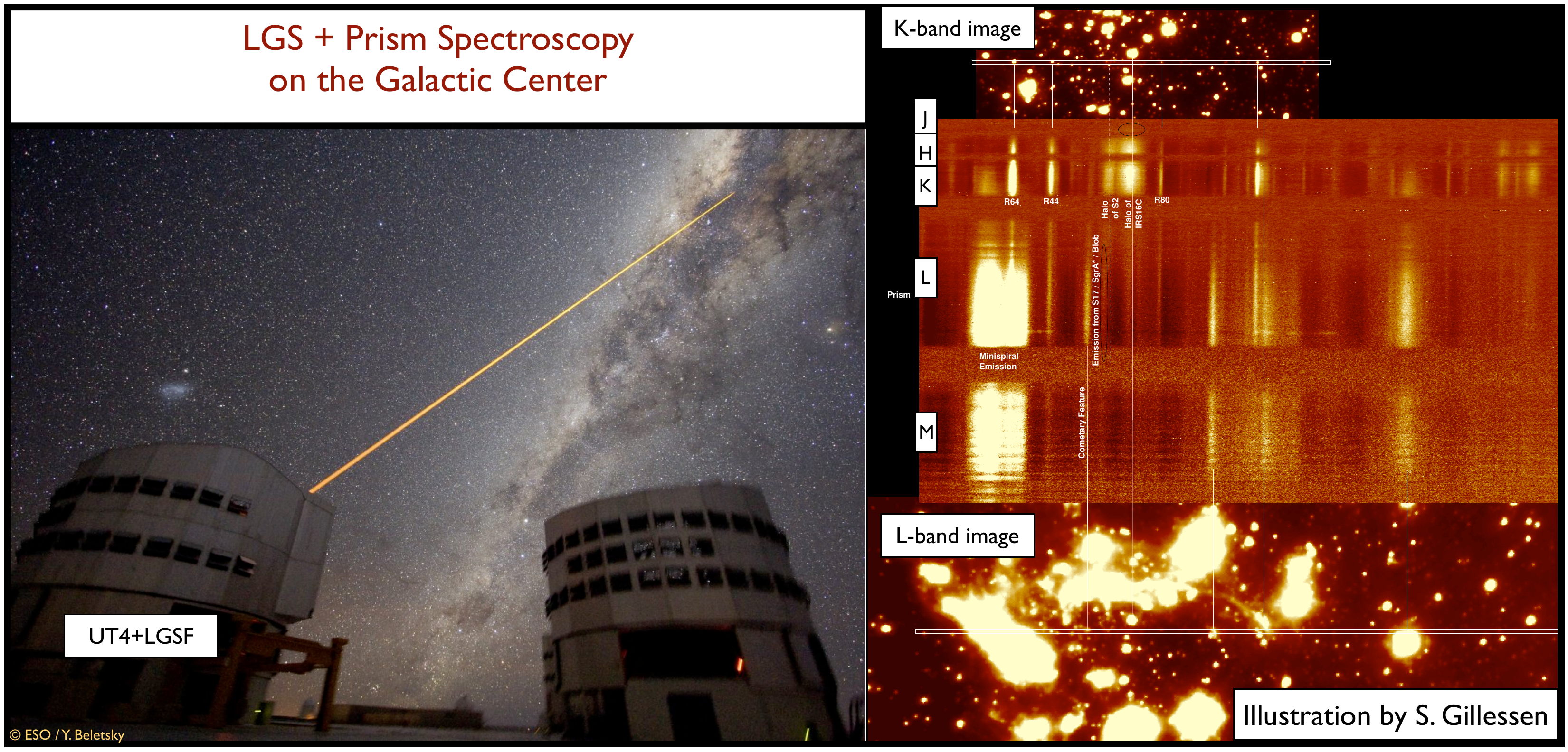}
\caption{Illustration of the potential of the prism spectroscopy as simultaneous photometry from J to M bands. Thanks to the use of the LGS the whole NIR spectrum from   1 to over 4 $\mu$m can be used for the scientific detector.
\label{fig:prism} 
}
\ec
\end{figure}

To illustrate one powerful technique associated to the operations with LGS. The well known galactic center group lead by R. Genzel has been recently awarded time on \naco to use prism spectroscopy (not currently offered) associated with LGS AO correction.
The great potential of this association of modes is illustrated in Figure \ref{fig:prism}. Although the weather prevented the team from getting any scientifically useful data, images in Lp and Ks bands were acquired and matched spatially to a spectrum acquire with the prism at very low spectral resolution. The idea is to carry out photometry from 1 to 4 $\mu m$ simultaneously, sending all the IR light to CONICA provided NAOS is correcting the wavefront via the LGS. A rough but sufficient spectral calibration was made possible thanks to imaging the slit through every possible narrow and intermediate-band filters inside CONICA and extrapolating the dispersion solution across the detector array. Though spectral lines are not properly resolved, one can associate one pixel coordinates to one wavelength. No flare occurred during this preliminary run.

\subsection{LGS: improved performances} 
\label{sec:vilu}

\begin{figure}[h!]
\bc
\includegraphics[width=0.85\textwidth]{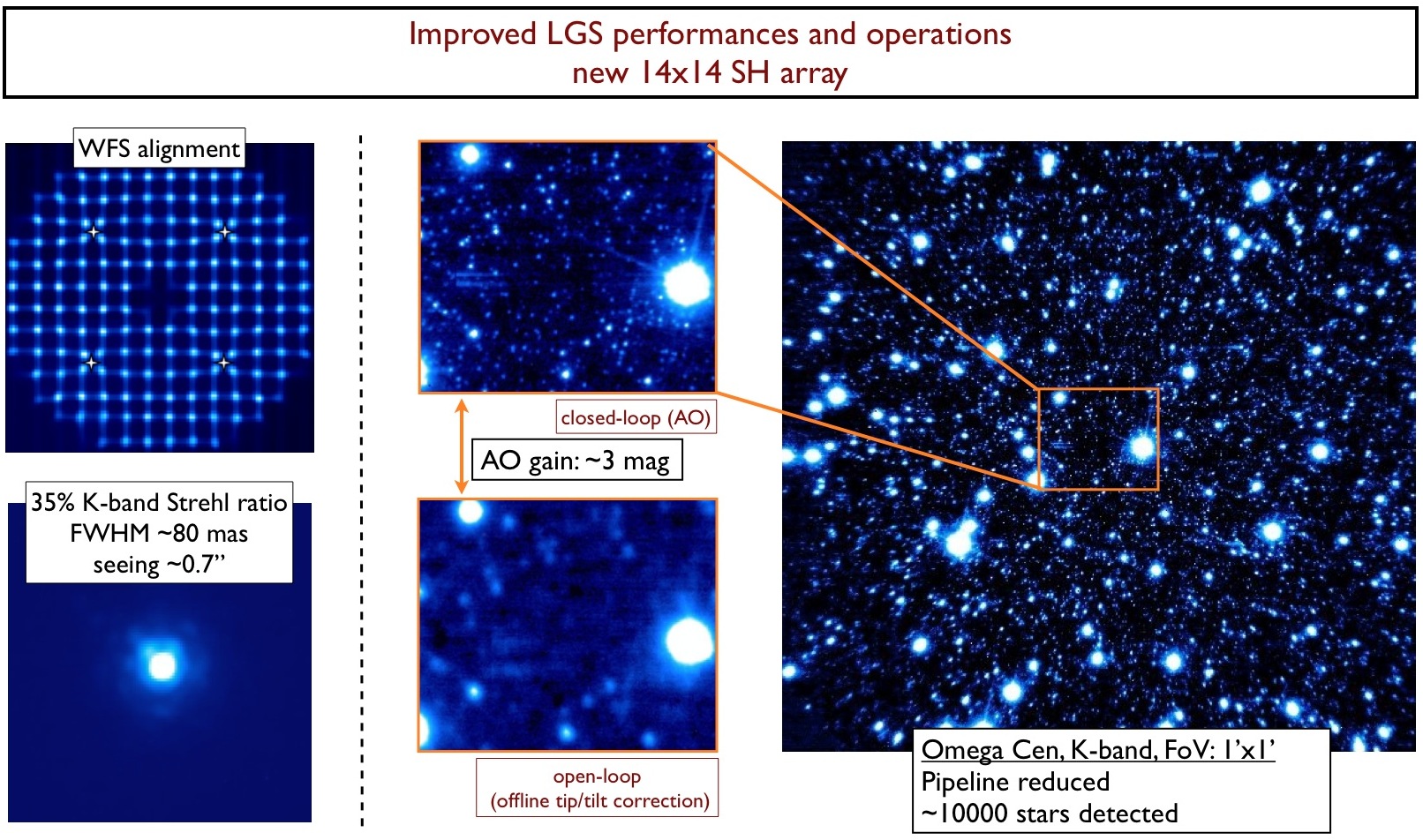}
\caption{Improved performances of the LGS with \naco due to the new 14$\times$14 lenslet array visible on the top left of the Figure (four actuators are poked for alignment purposes). The image at the bottom left corner shows a promising 35\% K-band strehl ratio (diffraction limited core) obtained with the LGS. The center and right images illustrate the potential of wide field correction with a  1\arcmin$\times$1\arcmin field centered on the star cluster Omega Centauri. 
\label{fig:14wfs} 
}
\ec
\end{figure}

Since the PARSEC\cite{rabien2002} laser guide star facility \cite{kasper2004, bonaccini2006} (LGSF) was introduced at the VLT in Nov 2007, it has gradually improved to lately become a robust and reliable system, scheduled approximately 8 to 10 nights per month with \naco or SINFONI. 
However, when feeding NACO, only the 7$\times$7 lenselet array could be used under median atmospheric conditions due to the LGS spot size and brightness with respect to the field of view and speed of the WFS. Therefore, the performances with LGS were significantly lower than with a natural guide star (NGS) even using the object as tip/tilt (TT) star. The expected 20\% Ks-band strehl ratio (SR) loss due to the cone effect (LGS at finite distance) was never reached. Considering a decent Ks-band SR of 50\% achievable with NGS, one could expect to get around 35 to 40\% using the LGS, which corresponds to a V$\sim$12 at nominal laser power. In February 2010, a new WFS array with 14$\times$14 lenslets was installed by a team composed by the original NAOS engineers of LAOG and ESO staff. Each of the sub-pupils now sees a field of view twice as large as the NGS 14$\times$14 array (4.6\arcsec instead of 2.3\arcsec) hence the LGS spot centroids (local wavefront slopes) are well calculated and the performances are improved at constant laser power and atmospheric conditions. The Figure \ref{fig:14wfs} shows an image of the new array, a K-band image featuring a 35\% strehl ratio obtained with a good but not exceptional seeing, and a 1\arcmin$\times$1\arcmin corrected field centered on the star cluster Omega Centauri\cite{kasper2010lgs}. The gain in sensitivity beween open-loop to closed-loop is tremendous (at least 3 magnitudes). On this image, anisoplanestism errors are seen on the edges of the field. However, they are small because the seeing, coherence time and isoplanetic patch were good, better than average on Paranal. Another reason to the somewhat wide field AO correction obtained here is the fact that LGS and TT star aren't perfectly superimposed. The TT star is the brightest star close to the center of the field. If the LGS is slightly off-axis,  not a single star will benefit from the optimal high order AO correction and TT correction however, a larger area will benefit from both.

The new lenslet array is now offered transparently to the user, when the atmospheric conditions and laser power allow its use.
Typically one needs a coherence time above 2.5 ms and a continuous laser power (off to the sky) of over 4W.
Since April 2010, the Seeing Enhancer (SE) mode was introduced to the community, allowing operations without TT correction and therefore full sky coverage. SE mode brings \naco to the extragalactic community. Even if the seeing is only improved by a factor two along the line of sight, the sensitivity is multiplied by a factor four and hence, high-redshift galaxies can be imaged in the NIR.

\subsection{Cube mode: powerful and popular} 
\label{sec:cube}

\begin{figure}[h!]
\centering
\includegraphics[width=0.85\textwidth]{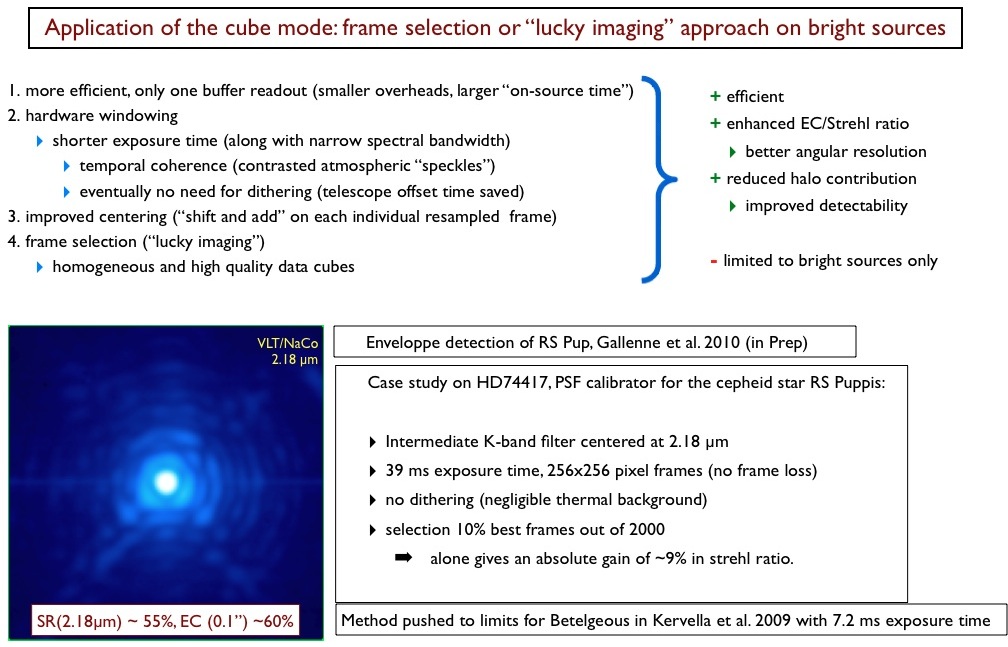}
\caption{
Cube mode applied to AO correction enhancement on bright sources.\protect\cite{kervella2009}.
\label{fig:cube} 
}
\end{figure}

Cube mode (sometimes called "burst mode") has been implemented in 2008. It is now widely used because of the reasons exposed in Figure \ref{fig:cube} and not only on bright sources because  it is proven to be more  efficient in time than the classical readout modes. Indeed, when cube mode is enabled, each frame (DIT\footnote[7]{DIT: Discrete Integration Time}) is transfered to a buffer which is read only at the very end of the sequence of DIT (NDIT). It is more efficient in terms of overheads (less down time at the VLT) and allows to have more control on the images via postprocessing. Since the best frames can be selected (the least affected by the atmospheric turbulence or the best AO corrected ones), centered one by one at sub-pixel accuracy, one can get enhanced final  image quality, improved detection limits, etc.

Today, the cube mode is not fully supported by the \naco pipeline which uses the mean of all images (just as if no cube mode was used) to produced the final stacked image. When this is still very useful for a quick look on the data quality, the expert user is encouraged to develop his/her own routines to fully exploit the potential of the mode. In the case of pupil tracking mode applied for angular differential imaging (ADI), a large image smearing is observed close to zenith due to the fast rotation of the field. The cube mode is then essential to recenter and derotate each frame, preserving the contrast, hence the angular resolution. 




\section{NEW OPERATION MODES} 
\label{sec:title}

In this section, new operation modes are described briefly with their performances and parameter spaces. The first one is about coronagraphy and direct imaging of extrasolar planets. The second one brings back older high angular resolution techniques that were used when simple AO was still a technological challenge. These new modes are designed for "niche science" as they are only useful under certain assumptions and/or conditions.

\subsection{APP: Apodizing Phase Plate coronagraph} 
\label{sec:app}

\begin{figure}[h!]
\centering
\includegraphics[width=0.75\textwidth]{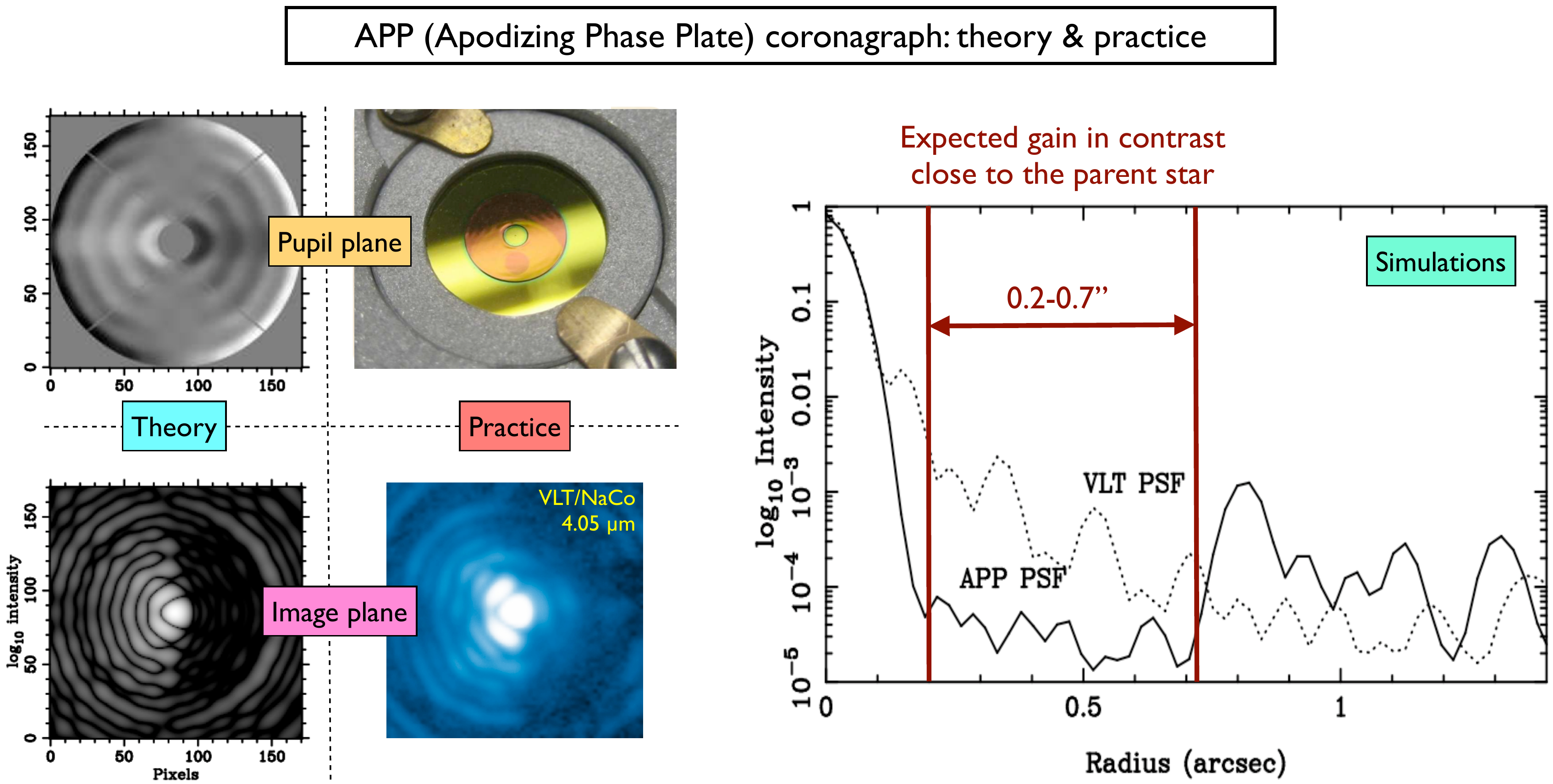}
\caption{
APP concept, realization and expected gain. A comparison of the theoretical PSF with respect to the real PSF obtained with \naco is also shown on the left. 
\label{fig:app1} 
}
\end{figure}

The Apodizing Phase Plate (APP) coronagraph \cite{codona2006, codona2007, kenworthy2007}was installed inside CONICA in November 2009 and successfully commissioned  \cite{kenworthy2010spie, app_com2010} in April 2010. It is offered for the P86 period starting in October this year. It consist of a plate in the pupil plane which generates certain optical path differences  causing the PSF in the focal plane to be half-obscured as shown on Figure \ref{fig:app1}. 

\begin{figure}[h!]
\bc
\includegraphics[width=0.75\textwidth]{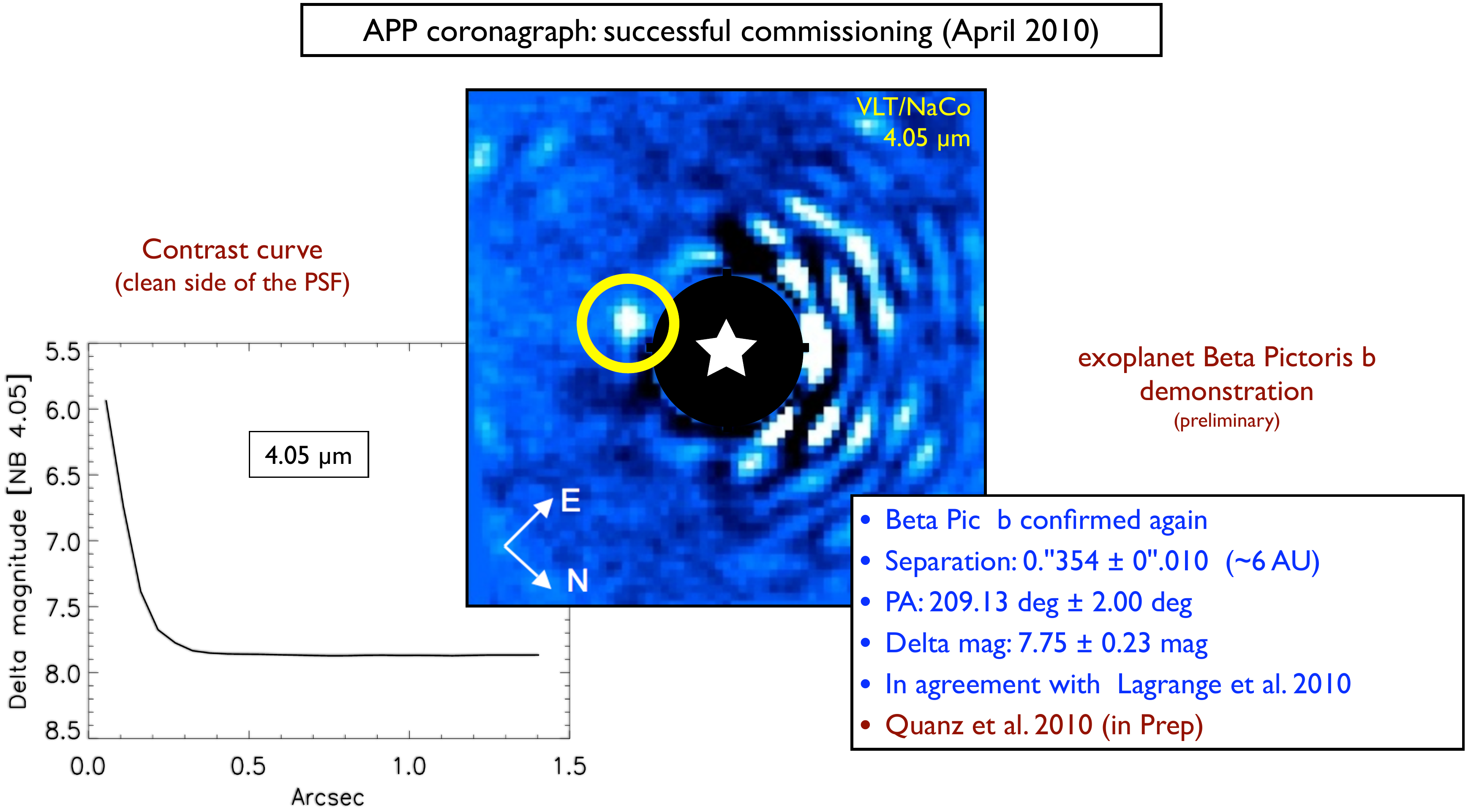}
\caption{
Main results of the \naco APP commissioning. The plot shown on the left is the contrast curve measured on the clean part of the PSF on an unresolved star under good but not exceptional atmospheric conditions (seeing around 0.7"). In the center is shown the reduced Beta Pictoris image obtained after PSF subtraction, recentering and derotation. The exoplanet Beta Pictoris b is clearly enhanced by the APP and all parameters (brightness ratio, position angle, separation) are in total agreement with the recent Science paper by Lagrange {\sl et al.}\protect\cite{lagrange2010}.
\label{fig:app2} 
}
\ec
\end{figure}

Designed for the L-band, it provides an improved detection limit from 1.8 to 7 $\lambda/D$ at 4.08 $\mu$ m (Figure \ref{fig:app2}, an optimal wavelength for direct imaging of cool extrasolar planets. The contrast curve at the right of Figure \ref{fig:app1} is flat on the obscured side as diffraction rings are suppressed. This translates to nearly a 8 magnitude difference detection capability from 0.2 to 0.7" arcsecond distance from the parent star ($\Delta$mag of 10 is achieved in L'-band).

The APP affects any point of the field and therefore does not require any special alignement like the classical focal plane masks. APP data is thus processed exactly the same way as classical imaging data. The APP performances are enhanced when combined to cube mode and pupil tracking.

\subsection{noAO: open-loop "speckle" mode} 
\label{sec:noao}

Technical time was spent in january 2010 to investigate the performances of open-loop imaging with \naco. One could think speckle imaging techniques are outdated, especially with an AO equipped 8-meter telescope. However We have reasons to believe that for certain applications and under mediocre atmospheric conditions, the noAO mode could be useful, particularly for astrometric purposes.

\begin{figure}[h!]
\centering
\includegraphics[width=0.6\textwidth]{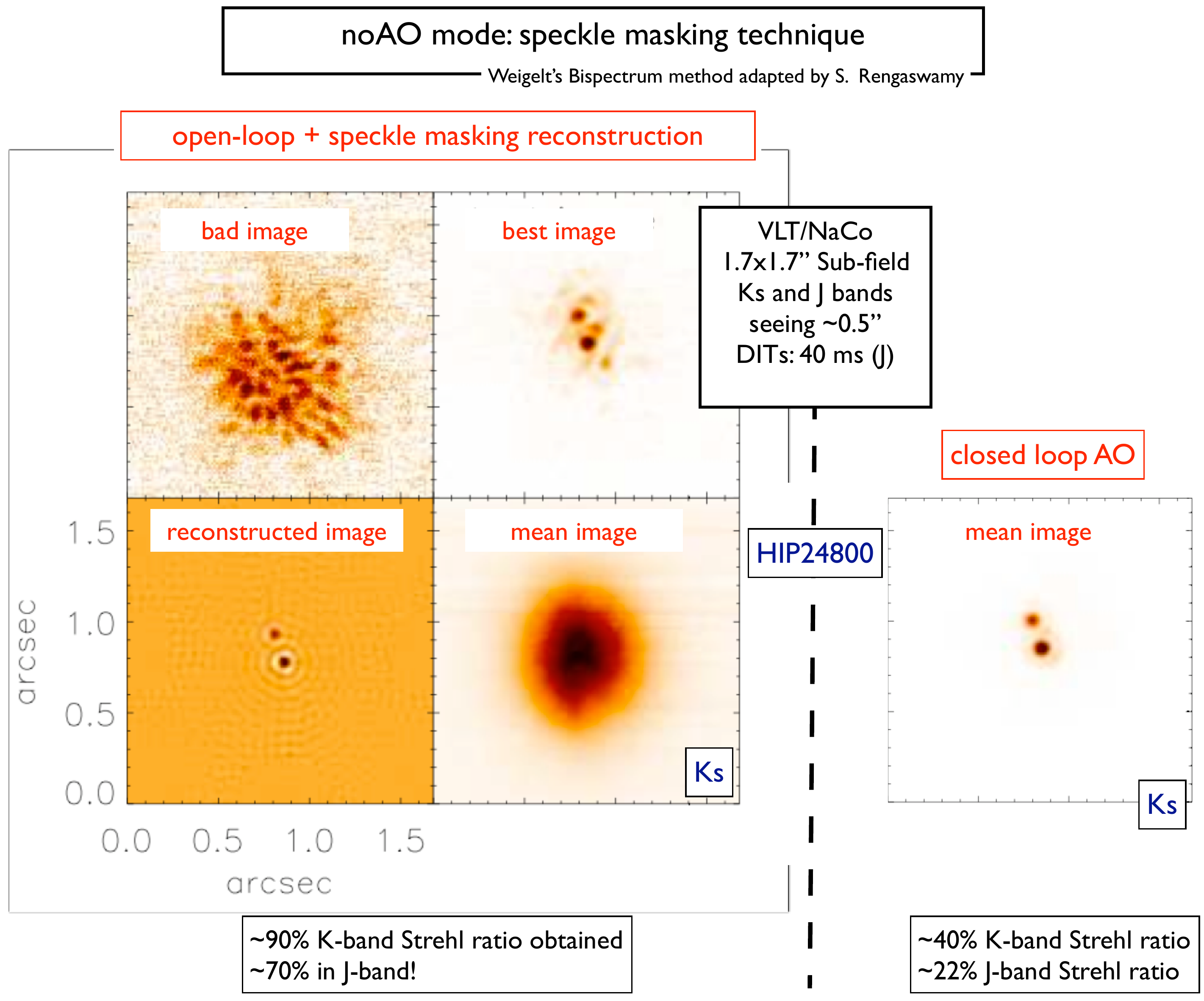}
\caption{
Some results of the speckle masking experiment \cite{sridhar2010speckle} applied on a 0.16\arcsec separation binary star. Comparison between the reconstructed intensity map (bottom left) and the closed-loop AO image (bottom right) is given. Seeing conditions were excellent for this test.
\label{fig:masking} 
}
\end{figure}

Two different approaches were then applied for the data analysis.
The first one (Figure \ref{fig:masking}) makes use of Weigelt's bispectrum method (here after called 
“speckle masking”). This project led by Sridharan Rengaswamy \cite{sridhar2010speckle} gives 
promising results for small fields of view, strehl ratios of the order of 60-90\% (J to K bands), 
and the ability to separate very close companions. On rather bright objects the reconstructed intensity maps
surpass the AO corrected images - taken the same night- though the absolute photometric calibration is less precise (of the order of 10\%). It should also be applicable to extended objects and represents a good alternative to AO for the 1-1.2 $\mu$m wavelength range or for any wavelength when the seeing conditions are too bad for AO.

\begin{figure}[h!]
\centering
\includegraphics[width=0.8\textwidth]{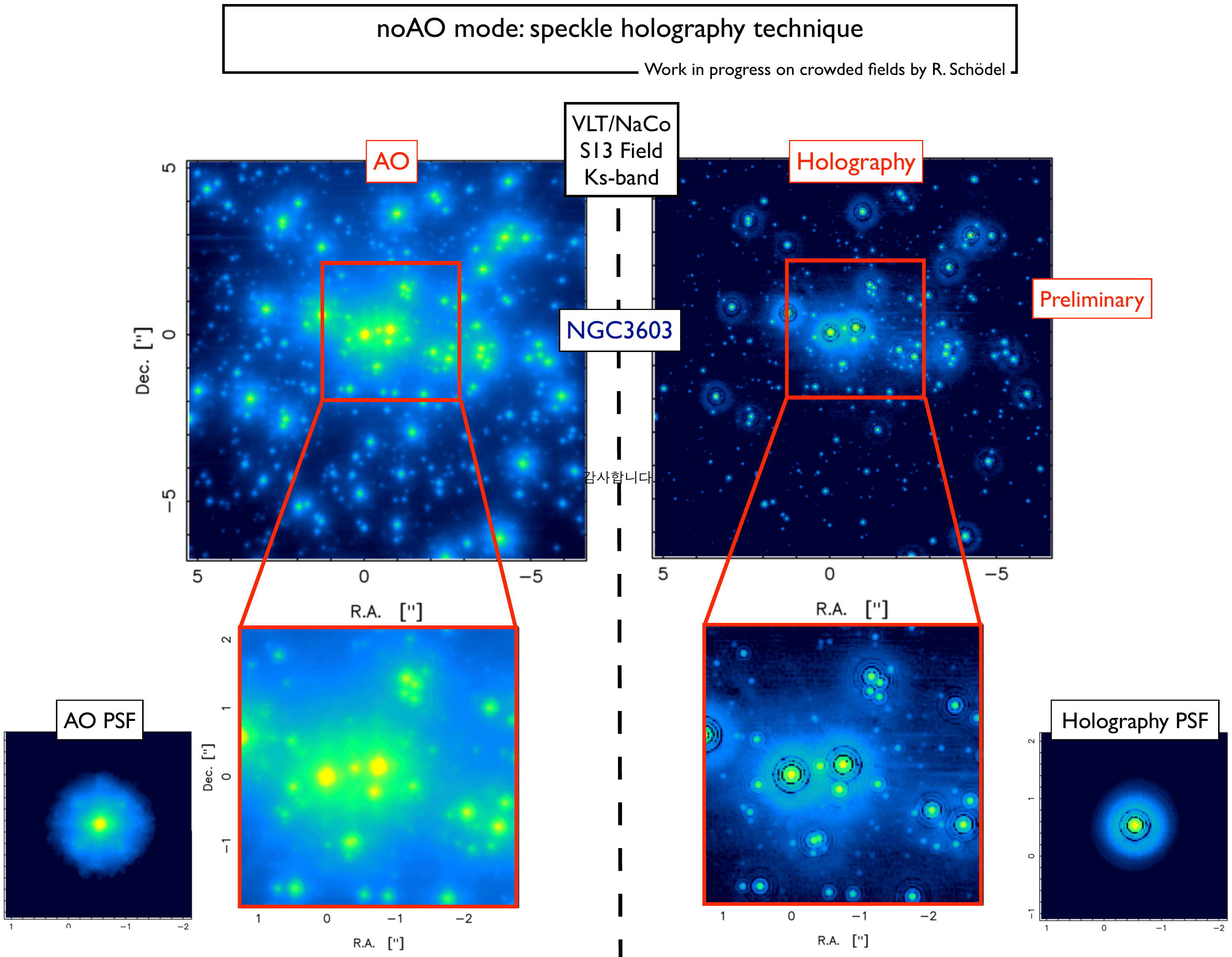}
\caption{
Speckle holography technique applied to the NGC3603 cluster. The holography image on the right is nearly as deep as the AO-corrected one on the left (the 3-$\sigma$ detection limit is $\sim$$K_{s} = 18$ in both cases). However it is advantageously more suitable for astrometry as every single star of the field appears in its true position, unaffected by AO-induced distorsion effects. The holography PSF itself is very clean, diffraction limited with a K-band strehl reaching 65\%, about 2.5 times superior to the closed-loop PSF which suffers an obvious waffle-mode pattern.
\label{fig:holo} 
}
\end{figure}

For the second approach(Figure \ref{fig:holo}), Rainer Sch\"{o}del is revisiting the speckle holography technique. 
The reference star for image reconstruction can be chosen {\sl a posteriori} and it is possible to use several reference stars simultaneously.  This opens the possibility to deal with effects of anisoplaneticity. The diffraction limit is reached over a larger field and hence this method is particularly suited for precise astrometry.

Just as cube mode, the noAO mode is an "expert mode" and no pipeline support will be provided. A quick look at the last image of each cube, seeing limited, ensures of the presence of the object onto the CONICA subfield.

\subsection{Sparse Aperture Masking (SAM) interferometry and pupil tracking} 
\label{sec:noao}

A paper by Tuthill {\sl et al.} was presented at this conference \cite{tuthill2010spie} about the SAM technique. Several aperture masks can be placed in a pupil plane inside CONICA and the mode is getting closer to its maturity.

As a by product of the SAM mode implementation, \naco can now also be operated in pupil tracking (PT) in classical imaging modes or coronagraphy, providing a somewhat stabilized PSF and a rotating field of view allowing angular differential imaging (ADI)\cite{marois2006}.  ADI is not offered as a mode {\sl per se} but it has become recurrently requested because it allows a clean PSF subtraction without the need of observing a reference star at the same paralactic angle. Of course, PT at a nasmyth focus can produce undesirable effects such as image drifts when the object passes close to zenith and the field rotation is fast.

\section{IDEAS} 
\label{sec:ideas}

There are a number of ideas to keep upgrading \naco. They arise from various teams, consortia and sometimes are just obvious or natural because some new concepts need an 8-meter class telescope to be tested and to provide their best. Of course, none of these are officially planned at the moment and they could only be programed if \naco were to stay longer at the VLT nasmyth focus. In this paper, we are only mentioning a few ideas, independently from any official or strategical standpoints. Here is a non-exhaustive list of ideas:

\bl
\item New SAM masks (e.g., an non-redundant anular mask).
\item A Vector Vortex Coronagraph\cite{mawet2009}(VVC): perhaps the most pertinent and feasible idea in the context of L-band imaging of extrasolar planets given the exciting results obtained at Mount Palomar\cite{serabyn2010} on the tripple system HR8799. A VVC working above 3 $\mu$m would have to be manufactured and inserted inside CONICA.
\item Fast CCD camera at visitor focus or in place of CONICA: speckle interferometry and lucky imaging in the visible.
\item SAM with pupil densification\cite{labeyrie1996}: the concept of hypertelescope could be tested on sky with a few tens of sub-pupils \protect\cite{patru2010spie}. Quite difficult to implement as two sets of microlenses would have to be aligned inside CONICA at low temperatures.
\item etc.
\el

\section{Conclusion} 
\label{sec:life}

\naco is one of the most requested ground-based instrument at the VLT and therefore in the world. It has generated original and well cited publications, some of which corresponding to ESO's cornerstone results. At nearly 10 years of extensive use, the question of maintaining \naco or replacing it arises while several second-generation VLT instruments are being manufactured.
\naco not only serves as test bench and survey tool for SPHERE but also allows to carry on programs in a different parameter space, notably thanks to L-band imaging/spectroscopy, IR wavefront sensing and LGS aided observations.
Lately, the need for a new 1-5 $\mu$m diffraction-limit imager for the VLT has been addressed \cite{ott2009vlt}
to fill the gap left by \naco once decommissioned. Nevertheless the \naco community is pushing hard to keep it on-sky another few periods.

\acknowledgments     
 
Julien Girard would like to acknowledge his predecessors as \naco Instrument Scientists  and/or AO specialists: Chris Lidman, Nancy Ageorges, Paola Amico, Emanuela Pompei. He thanks the whole IOT in particular  Lowell Tacconi-Garman who provided very useful feedback on the elaboration of this paper. He also thanks staff at ESO and LAOG at large who provided help on the implementation of the new modes of \naco, in particular  St\'ephane Marteau, Eric Stadler, Philippe Feautrier, Christian S\"onke, Patrick Rabou, etc.
Finally he thanks ESO Paranal management, Engineering, Science Operation Head Christophe Dumas, C\'edric Ledoux,  and IOT Coordinator Alain Smette for their valuable inputs and support.

\bibliography{../../../BIB/biblio}

\begin{thebibliography}{10}

\bibitem{brandner2002}
{Brandner}, W., {Rousset}, G., {Lenzen}, R., {Hubin}, N., {Lacombe}, F.,
  {Hofmann}, R., {Moorwood}, A., {Lagrange}, A., {Gendron}, E., {Hartung}, M.,
  {Puget}, P., {Ageorges}, N., {Biereichel}, P., {Bouy}, H., {Charton}, J.,
  {Dumont}, G., {Fusco}, T., {Jung}, Y., {Lehnert}, M., {Lizon}, J., {Monnet},
  G., {Mouillet}, D., {Moutou}, C., {Rabaud}, D., {R{\"o}hrle}, C., {Skole},
  S., {Spyromilio}, J., {Storz}, C., {Tacconi-Garman}, L., and {Zins}, G.,
  ``{NAOS+CONICA at YEPUN: first VLT adaptive optics system sees first
  light},'' {\em The Messenger} {\bf 107},  1--6 (Mar. 2002).

\bibitem{lenzen2003}
{Lenzen}, R., {Hartung}, M., {Brandner}, W., {Finger}, G., {Hubin}, N.~N.,
  {Lacombe}, F., {Lagrange}, A., {Lehnert}, M.~D., {Moorwood}, A.~F.~M., and
  {Mouillet}, D., ``{NAOS-CONICA first on sky results in a variety of observing
  modes},'' in [{\em Society of Photo-Optical Instrumentation Engineers (SPIE)
  Conference Series}{\nolinebreak\hspace{0.1em}]},  {M.~Iye \&
  A.~F.~M.~Moorwood}, ed., {\em Society of Photo-Optical Instrumentation
  Engineers (SPIE) Conference Series} {\bf 4841},  944--952 (Mar. 2003).

\bibitem{rousset2003}
{Rousset}, G., {Lacombe}, F., {Puget}, P., {Hubin}, N.~N., {Gendron}, E.,
  {Fusco}, T., {Arsenault}, R., {Charton}, J., {Feautrier}, P., {Gigan}, P.,
  {Kern}, P.~Y., {Lagrange}, A., {Madec}, P., {Mouillet}, D., {Rabaud}, D.,
  {Rabou}, P., {Stadler}, E., and {Zins}, G., ``{NAOS, the first AO system of
  the VLT: on-sky performance},'' in [{\em Society of Photo-Optical
  Instrumentation Engineers (SPIE) Conference
  Series}{\nolinebreak\hspace{0.1em}]},  {P.~L.~Wizinowich \&amp;
  D.~Bonaccini}, ed., {\em Society of Photo-Optical Instrumentation Engineers
  (SPIE) Conference Series} {\bf 4839},  140--149 (Feb. 2003).

\bibitem{schoedel2002}
{Sch{\"o}del}, R., {Ott}, T., {Genzel}, R., {Hofmann}, R., {Lehnert}, M.,
  {Eckart}, A., {Mouawad}, N., {Alexander}, T., {Reid}, M.~J., {Lenzen}, R.,
  {Hartung}, M., {Lacombe}, F., {Rouan}, D., {Gendron}, E., {Rousset}, G.,
  {Lagrange}, A., {Brandner}, W., {Ageorges}, N., {Lidman}, C., {Moorwood},
  A.~F.~M., {Spyromilio}, J., {Hubin}, N., and {Menten}, K.~M., ``{A star in a
  15.2-year orbit around the supermassive black hole at the centre of the Milky
  Way},'' {\em Nature} {\bf 419},  694--696 (Oct. 2002).

\bibitem{genzel2003}
{Genzel}, R., {Sch{\"o}del}, R., {Ott}, T., {Eisenhauer}, F., {Hofmann}, R.,
  {Lehnert}, M., {Eckart}, A., {Alexander}, T., {Sternberg}, A., {Lenzen}, R.,
  {Cl{\'e}net}, Y., {Lacombe}, F., {Rouan}, D., {Renzini}, A., and
  {Tacconi-Garman}, L.~E., ``{The Stellar Cusp around the Supermassive Black
  Hole in the Galactic Center},'' {\em ApJ} {\bf 594},  812--832 (Sept. 2003).

\bibitem{eckart2008}
{Eckart}, A., {Baganoff}, F.~K., {Zamaninasab}, M., {Morris}, M.~R.,
  {Sch{\"o}del}, R., {Meyer}, L., {Muzic}, K., {Bautz}, M.~W., {Brandt}, W.~N.,
  {Garmire}, G.~P., {Ricker}, G.~R., {Kunneriath}, D., {Straubmeier}, C.,
  {Duschl}, W., {Dovciak}, M., {Karas}, V., {Markoff}, S., {Najarro}, F.,
  {Mauerhan}, J., {Moultaka}, J., and {Zensus}, A., ``{Polarized NIR and X-ray
  flares from Sagittarius A*},'' {\em A\&A} {\bf 479},  625--639 (Mar. 2008).

\bibitem{murakawa2008}
{Murakawa}, K., {Preibisch}, T., {Kraus}, S., {Ageorges}, N., {Hofmann}, K.,
  {Ishii}, M., {Oya}, S., {Rosen}, A., {Schertl}, D., and {Weigelt}, G.,
  ``{VLT/NACO and Subaru/CIAO JHK-band high-resolution imaging polarimetry of
  the Herbig Be star R Monocerotis},'' {\em A\&A}~{\bf 488},  L75--L78 (Sept.
  2008).

\bibitem{sicardy2006}
{Sicardy}, B., {Bellucci}, A., {Gendron}, E., {Lacombe}, F., {Lacour}, S.,
  {Lecacheux}, J., {Lellouch}, E., {Renner}, S., {Pau}, S., {Roques}, F.,
  {Widemann}, T., {Colas}, F., {Vachier}, F., {Martins}, R.~V., {Ageorges}, N.,
  {Hainaut}, O., {Marco}, O., {Beisker}, W., {Hummel}, E., {Feinstein}, C.,
  {Levato}, H., {Maury}, A., {Frappa}, E., {Gaillard}, B., {Lavayssi{\`e}re},
  M., {di Sora}, M., {Mallia}, F., {Masi}, G., {Behrend}, R., {Carrier}, F.,
  {Mousis}, O., {Rousselot}, P., {Alvarez-Candal}, A., {Lazzaro}, D., {Veiga},
  C., {Andrei}, A.~H., {Assafin}, M., {da Silva Neto}, D.~N., {Jacques}, C.,
  {Pimentel}, E., {Weaver}, D., {Lecampion}, J., {Doncel}, F., {Momiyama}, T.,
  and {Tancredi}, G., ``{Charon's size and an upper limit on its atmosphere
  from a stellar occultation},'' {\em Nature} {\bf 439},  52--54 (Jan. 2006).

\bibitem{chauvin2004}
{Chauvin}, G., {Lagrange}, A., {Dumas}, C., {Zuckerman}, B., {Mouillet}, D.,
  {Song}, I., {Beuzit}, J., and {Lowrance}, P., ``{A giant planet candidate
  near a young brown dwarf. Direct VLT/NACO observations using IR wavefront
  sensing},'' {\em A\&A} {\bf 425},  L29--L32 (Oct. 2004).

\bibitem{neuhauser2005}
{Neuh{\"a}user}, R., {Guenther}, E.~W., {Wuchterl}, G., {Mugrauer}, M.,
  {Bedalov}, A., and {Hauschildt}, P.~H., ``{Evidence for a co-moving
  sub-stellar companion of GQ Lup},'' {\em A\&A} {\bf 435},  L13--L16 (May
  2005).

\bibitem{kasper2005}
{Kasper}, M., {Ageorges}, N., {Boccaletti}, A., {Brandner}, W., {Close}, L.~M.,
  {Davies}, R., {Finger}, G., {Genzel}, R., {Hartung}, M., {Kaufer}, A.,
  {Kellner}, S., {Hubin}, N., {Lenzen}, R., {Ludman}, C., {Monnet}, G.,
  {Moorwood}, A., {Ott}, T., {Riaud}, P., {Roser}, H., {Rouan}, D., and
  {Spyromilio}, J., ``{New observing modes of NACO},'' {\em The Messenger}~{\bf
  119},  11--+ (Mar. 2005).

\bibitem{rouan2000}
{Rouan}, D., {Riaud}, P., {Boccaletti}, A., {Cl{\'e}net}, Y., and {Labeyrie},
  A., ``{The Four-Quadrant Phase-Mask Coronagraph. I. Principle},'' {\em
  PASP} {\bf 112},  1479--1486 (Nov. 2000).

\bibitem{witzel2010}
{Witzel}, G., {Eckart}, A., {Buchholz}, R.~M., {Zamaninasab}, M., {Lenzen}, R.,
  {Sch\"{o}del}, R., {Araujo}, {Sabha}, N., {Bremer}, M., {Karas}, V.,
  {Straubmeier}, C., and {Muzic}, K., ``{The instrumental polarization of the
  Nasmyth focus polarimetric differential imager NAOS/CONICA (NACO) at the VLT,
  Implications for time resolved polarimetric measurements of Sgr A*},'' {\em
  A\&A} {\bf in prep} (2010).

\bibitem{kasper2009}
{Kasper}, M., {Amico}, P., {Pompei}, E., {Ageorges}, N., {Apai}, D.,
  {Argomedo}, J., {Kornweibel}, N., and {Lidman}, C., ``{Direct Imaging of
  Exoplanets and Brown Dwarfs with the VLT: NACO Pupil-stabilised Lyot
  Coronagraphy at 4 micron},'' {\em The Messenger} {\bf 137},  8--13 (Sept.
  2009).

\bibitem{janson2010}
{Janson}, M., {Bergfors}, C., {Goto}, M., {Brandner}, W., and {Lafreni{\`e}re},
  D., ``{Spatially Resolved Spectroscopy of the Exoplanet HR 8799 c},'' {\em
  ApJL}~{\bf 710},  L35--L38 (Feb. 2010).

\bibitem{dobrzycka2004}
{Dobrzycka}, D., {Hummel}, W., {Lidman}, C., {Ageorges}, N., {Marco}, O., and
  {Jung}, Y., ``{Quality control of VLT NACO data},'' in [{\em Society of
  Photo-Optical Instrumentation Engineers (SPIE) Conference
  Series}{\nolinebreak\hspace{0.1em}]},  {P.~J.~Quinn \& A.~Bridger}, ed., {\em
  Presented at the Society of Photo-Optical Instrumentation Engineers (SPIE)
  Conference} {\bf 5493},  579--583 (Sept. 2004).

\bibitem{rabien2002}
{Rabien}, S., {Davies}, R.~I., {Ott}, T., {Hippler}, S., and {Neumann}, U.,
  ``{PARSEC: the laser for the VLT},'' in [{\em Society of Photo-Optical
  Instrumentation Engineers (SPIE) Conference
  Series}{\nolinebreak\hspace{0.1em}]},  {R.~K.~Tyson, D.~Bonaccini, \&
  M.~C.~Roggemann}, ed., {\em Presented at the Society of Photo-Optical
  Instrumentation Engineers (SPIE) Conference} {\bf 4494},  325--335 (Feb.
  2002).

\bibitem{kasper2004}
{Kasper}, M.~E., {Charton}, J., {Delabre}, B., {Donaldson}, R., {Fedrigo}, E.,
  {Hess}, G., {Hubin}, N.~N., {Lizon}, J., {Nylund}, M., {Soenke}, C., and
  {Zins}, G., ``{LGS implementation for NAOS},'' in [{\em Society of
  Photo-Optical Instrumentation Engineers (SPIE) Conference
  Series}{\nolinebreak\hspace{0.1em}]},  {D.~Bonaccini Calia, B.~L.~Ellerbroek,
  \&amp; R.~Ragazzoni}, ed., {\em Presented at the Society of Photo-Optical
  Instrumentation Engineers (SPIE) Conference} {\bf 5490},  1071--1078 (Oct.
  2004).

\bibitem{bonaccini2006}
{Bonaccini Calia}, D., {Allaert}, E., {Alvarez}, J.~L., {Araujo Hauck}, C.,
  {Avila}, G., {Bendek}, E., {Buzzoni}, B., {Comin}, M., {Cullum}, M.,
  {Davies}, R., {Dimmler}, M., {Guidolin}, I., {Hackenberg}, W., {Hippler}, S.,
  {Kellner}, S., {van Kesteren}, A., {Koch}, F., {Neumann}, U., {Ott}, T.,
  {Popovic}, D., {Pedichini}, F., {Quattri}, M., {Quentin}, J., {Rabien}, S.,
  {Silber}, A., and {Tapia}, M., ``{First light of the ESO Laser Guide Star
  Facility},'' in [{\em Society of Photo-Optical Instrumentation Engineers
  (SPIE) Conference Series}{\nolinebreak\hspace{0.1em}]},  {\em Presented at
  the Society of Photo-Optical Instrumentation Engineers (SPIE) Conference}
  {\bf 6272} (July 2006).

\bibitem{kasper2010lgs}
{Kasper}, M., {Zins}, G., {Feautrier}, P., {O'Neal}, J., {Michaud}, L.,
  {Rabou}, P., {Stadler}, E., {Charton}, J., {Cumani}, C., {Delboulbe}, A.,
  {Geimer}, C., {Gillet}, G., {Girard}, J., {Huerta}, N., {Kern}, P., {Lizon},
  J., {Lucuix}, C., {Mouillet}, D., {Moulin}, T., {Rochat}, S., and
  {S{\"o}nke}, C., ``{A New Lenslet Array for the NACO Laser Guide Star
  Wavefront Sensor},'' {\em The Messenger}~{\bf 140},  8--9 (June 2010).

\bibitem{kervella2009}
{Kervella}, P., {Verhoelst}, T., {Ridgway}, S.~T., {Perrin}, G., {Lacour}, S.,
  {Cami}, J., and {Haubois}, X., ``{The close circumstellar environment of
  Betelgeuse. Adaptive optics spectro-imaging in the near-IR with VLT/NACO},''
  {\em A\&A}~{\bf 504},  115--125 (Sept. 2009).

\bibitem{codona2006}
{Codona}, J.~L., {Kenworthy}, M.~A., {Hinz}, P.~M., {Angel}, J.~R.~P., and
  {Woolf}, N.~J., ``{A high-contrast coronagraph for the MMT using phase
  apodization: design and observations at 5 microns and 2 {$\lambda$}/D
  radius},'' in [{\em Society of Photo-Optical Instrumentation Engineers (SPIE)
  Conference Series}{\nolinebreak\hspace{0.1em}]},  {\em Society of
  Photo-Optical Instrumentation Engineers (SPIE) Conference Series} {\bf 6269}
  (July 2006).

\bibitem{codona2007}
{Codona}, J., ``{Phase Apodization Coronagraphy},'' in [{\em In the Spirit of
  Bernard Lyot: The Direct Detection of Planets and Circumstellar Disks in the
  21st Century}{\nolinebreak\hspace{0.1em}]},  {Kalas}, P., ed.,  24--+ (June
  2007).

\bibitem{kenworthy2007}
{Kenworthy}, M.~A., {Codona}, J.~L., {Hinz}, P.~M., {Angel}, J.~R.~P.,
  {Heinze}, A., and {Sivanandam}, S., ``{First On-Sky High-Contrast Imaging
  with an Apodizing Phase Plate},'' {\em ApJ}~{\bf 660},  762--769 (May 2007).

\bibitem{kenworthy2010spie}
{Kenworthy}, M., {Meyer}, M., {Quanz}, S.~P., {Kasper}, M., {Lenzen}, R.,
  {Codona}, J., {Girard}, J. H.~V., and {Hinz}, P., ``{An Apodizing Phase Plate
  Coronagraph for the VLT},'' in [{\em {Ground-based and Airborne
  Instrumentation for Astronomy III, Proceedings of the
  SPIE}}{\nolinebreak\hspace{0.1em}]},  {McLean}, I.~S., K., R.~S., and
  {Takami}, H., eds., {\em Astronomical Instrumentation} {\bf 7735}, Society of
  Photo-Optical Instrumentation Engineers (SPIE) Conference Series (July 2010).

\bibitem{app_com2010}
{Meyer}, M.~R., {Quanz}, S.~P., {Kenworthy}, M., {Kasper}, M., {Lenzen}, R.,
  and {Girard}, J.~H., ``Naco app commissioning report,'' tech. rep., ETH
  Zurich, ESO, MPIA, Leiden Observatory (2010).

\bibitem{lagrange2010sci}
{Lagrange}, A., {Bonnefoy}, M., {Chauvin}, G., {Apai}, D., {Ehrenreich}, D.,
  {Boccaletti}, A., {Gratadour}, D., {Rouan}, D., {Mouillet}, D., {Lacour}, S.,
  and {Kasper}, M., ``{A Giant Planet Imaged in the Disk of the Young Star
  {$\beta$} Pictoris},'' {\em Science}~{\bf 329},  57-- (July 2010).

\bibitem{sridhar2010speckle}
{Rengaswamy}, S., {Girard}, J. H.~V., and {Montagnier}, G., ``{Speckle imaging
  with the SOAR and the very large telescopes},'' in [{\em {Optical and
  Infrared Interferometry II, Proceedings of the
  SPIE}}{\nolinebreak\hspace{0.1em}]},  {Danchi}, W.~C., {Delplancke}, F., and
  {Rajagopal}, J.~K., eds., {\em Astronomical Instrumentation} {\bf 7734},
  Society of Photo-Optical Instrumentation Engineers (SPIE) Conference Series
  (July 2010).

\bibitem{tuthill2010spie}
{Tuthill}, P., {Lacour}, S., {Amico}, P., {Ireland}, M., {Norris}, B.,
  {Stewart}, P., {Evans}, T., {Kraus}, A., {Lidman}, C., {Pompei}, E., and
  {Kornweibel}, N., ``{Sparse Aperture Masking (SAM) at NAOS/CONICA on the
  VLT},'' {\em ArXiv e-prints}~{\bf 7735} (June 2010).

\bibitem{marois2006}
{Marois}, C., {Lafreni{\`e}re}, D., {Doyon}, R., {Macintosh}, B., and {Nadeau},
  D., ``{Angular Differential Imaging: A Powerful High-Contrast Imaging
  Technique},'' {\em ApJ}~{\bf 641},  556--564 (Apr. 2006).

\bibitem{mawet2009}
{Mawet}, D., {Serabyn}, E., {Liewer}, K., {Hanot}, C., {McEldowney}, S.,
  {Shemo}, D., and {O'Brien}, N., ``{Optical Vectorial Vortex Coronagraphs
  using Liquid Crystal Polymers: theory, manufacturing and laboratory
  demonstration},'' {\em Optics Express}~{\bf 17},  1902--1918 (Feb. 2009).

\bibitem{serabyn2010}
{Serabyn}, E., {Mawet}, D., and {Burruss}, R., ``{An image of an exoplanet
  separated by two diffraction beamwidths from a star},'' {\em Nature}~{\bf 464},
   1018--1020 (Apr. 2010).

\bibitem{labeyrie1996}
{Labeyrie}, A., ``{Resolved imaging of extra-solar planets with future 10-100km
  optical interferometric arrays.},'' {\em A\&As}~{\bf 118},  517--524 (Sept.
  1996).

\bibitem{patru2010spie}
{Patru}, F., {Mourard}, D., {Tarmoul}, N., and {Chiavassa}, D., ``{observation
  preparation and data reduction, its community remains limited. Direct imaging
  with a hypertelescope: array confi guration versus science cases},'' in [{\em
  {Optical and Infrared Interferometry II, Proceedings of the
  SPIE}}{\nolinebreak\hspace{0.1em}]},  {Danchi}, W.~C., {Delplancke}, F., and
  {Rajagopal}, J.~K., eds., {\em Astronomical Instrumentation} {\bf 7734},
  Society of Photo-Optical Instrumentation Engineers (SPIE) Conference Series
  (July 2010).

\bibitem{ott2009vlt}
{Ott}, T., {Davies}, R., {Eisenhauer}, F., {Genzel}, R., {Hofmann}, R., and
  {Gillessen}, S., ``{The Need for a General Purpose Diffraction Limited Imager
  at the VLT},'' {\em Science with the VLT in the ELT
  Era},  {A.~Moorwood}, ed.,  481--484 (2009).

\end{thebibliography}
\bibliographystyle{spiebib}   

\end{document}